\documentclass{aa} 
\usepackage{graphicx}
\usepackage{lscape}
\begin{document}        


\def\ra{\rightarrow} \def\deg{\ifmmode^\circ\else$^\circ$\fi}
\def\kms{km\thinspace s$^{-1}$}
\def\msun{M$_{\odot}$}
\def\lsun{L$_{\odot}$}
\def\h13co+{H$^{13}$CO$^+$}

\title{The Youngest Stellar clusters}
\subtitle{Clusters associated with massive protostellar candidates} 

\author{ M.  S.  N. Kumar\inst{1}, E. Keto\inst{2} \and  E. Clerkin\inst{1},\inst{3} }

\institute{Centro de Astrof\'{\i}sica da Universidade do Porto, Rua
das Estrelas, 7150-462 Porto, Portugal\\ e-mail:nanda@astro.up.pt,\and Harvard Smithsonian Center for Astrophysics, 60 Garden Street, Cambridge, Massachusetts, USA \and Present address: Shanghai Institute of Applied Maths and Mechanics, Shanghai University, Shanghai 200072, China.}

\offprints{M.  S.  Nanda Kumar} 
\date{Received:  ; accepted:  }
\authorrunning{Kumar, Keto \& Clerkin}  
\titlerunning{2MASS study of precursors to massive stars}

\abstract {
We report  on the  identification of 54   embedded clusters around 217
massive   protostellar candidates   of which    34 clusters  are   new
detections.  The embedded  clusters are identified as stellar  surface
density  enhancements  in the 2  $\mu$m  All Sky  Survey (2MASS) data.
Because the  clusters are all associated  with  massive stars in their
earliest evolutionary stage,  the clusters should  also be in an early
stage of evolution.   Thus  the properties of  these   clusters should
reflect  properties associated with their  formation rather than their
evolution.  For each cluster, we estimate  the mass, the morphological
type, the photometry and  extinction.  The clusters  in our  study, by
their  association with  massive   protostars  and massive   outflows,
reinstate the notion that massive stars begin to  form after the first
generation   of  low  mass   stars    have completed  their  accretion
phase. Further, the observed high gas densities and accretion rates at
the centers  of these clusters  is consistent with the hypothesis that
high mass stars form by continuing accretion onto low mass stars.

 }

\maketitle

\section{Introduction}

Embedded stellar clusters, those clusters that are still surrounded by
the  molecular clouds in  which they  formed, are  the youngest of the
stellar clusters. As such,   the embedded clusters are of   particular
interest    to understand which     properties of stellar clusters are
related  to  their  origins  and which   are derived  from  subsequent
evolution  (Elmegreen      et     al.~\cite{elmegreen00}; Lada     \&
Lada~\cite{ll03}).  For   example,    the  mass    segregation,   the
concentration of higher mass stars in the centers of clusters, that is
observed in many optically visible open  clusters is also seen in some
of the embedded clusters.  Because the embedded clusters are too young
to     have  undergone  significant   dynamical   evolution,  the mass
segregation must be a property of the process of star formation in the
clusters.  As another example, open clusters exhibit both hierarchical
and centrally   condensed morphological  types. Observations   of both
morphological types in  embedded clusters suggests that the morphology
of the clusters reflects  the morphology of  the clouds from which the
stars   formed  rather    than    the dynamical  evolution     of  the
cluster.  Finally, the distribution of  stellar masses in the embedded
clusters  ought to be   little  affected by  evolution and   therefore
closest to the initial mass function (IMF).

While the embedded clusters are thought to be among the youngest
clusters, if we were to identify a class of clusters in which star
formation, and therefore the formation of the cluster itself, were
just beginning, we would potentially be able to address some of the
questions as to the causes and origins of some of the cluster
properties. For example, observations of embedded clusters may suggest
star formation rather than dynamical evolution as a cause of mass
segregation, but there remains the question of the cause.  Is mass
segregation a result of the formation of massive stars by the
collisions and coalescence of lower mass stars because collisions will
be more common in the high stellar density in the center of a cluster
(Bonnell et al \cite{bonnell98}; Testi et al. \cite{testi99})? Is
mass segregation a result of the formation of massive stars by
continuing accretion onto existing low mass stars (Beech and
Mitalas~\cite{bm94}; Behrend \& Maeder~\cite{bm01}; Bernasconi and
Maeder~\cite{bm96}; Meynet and Maeder~\cite{mm00}, Keto
\cite{keto03}), a process that requires rapid accretion to overcome
radiation and thermal  pressure,  and therefore requires dense  gas as
would be  found in the center  of a dense molecular clouds? Similarly,
observations of actively  forming clusters  might potentially  address
the relationship     of   the  star formation  to      molecular cloud
structure. What  is the difference in  cloud structure leading  to the
hierarchical and  centrally condensed morphological types of clusters? 
Do  stars always form  first in the center of  a molecular cloud or in
gravitationally collapsing fragments throughout the cloud? Finally, if
the lower mass  stars form  first as suggested  by several  studies of
open  clusters (Herbig \cite{herbig62}, Stahler \cite{stahler85}), and
required  by the  theories of     massive star  formation either    by
coalescence or continuing accretion, then  observations of clusters in
formation may potentially address the origins of the IMF. Observations
of the  stellar mass  distribution  in actively  forming  clusters may
allow the  opportunity  to  see  mass   distributions that  are  still
evolving toward an IMF.

In this study, we test  a hypothesis that a  class of actively forming
clusters, the youngest subset  of the young  embedded clusters, may be
identified   by searching   for  stellar  clusters   around previously
identified   massive protostellar   candidates  in isolated  molecular
clouds.  These massive protostellar candidates, massive stars in their
earliest  stages  of formation,  have  been  identified  as a class of
luminous objects having specific IRAS colors  that are associated with
other indicators of massive star formation such as dense gas and dust,
water  masers, and    ultra-compact   HII (UCHII)  regions  (Palla  et
al.~\cite{palla91};                 Molinari                        et
al.~\cite{mol96},~\cite{mol98},~\cite{mol00};       Sridharan       et
al.~\cite{sri01};  Beuther et   al.\cite{beu02a})  and associated with
isolated  molecular  clouds.    These candidates  are   sources deeply
embedded in  their molecular clouds   and therefore probably represent
the first massive stars to form within the  clouds. If there were more
evolved  massive stars  in  these clouds,  then we  would expect their
winds, radiation pressure, and supernovae explosions  to have at least
partially cleared  the region  of  gas  and dust  revealing  perhaps a
classic  open cluster as  would typically be found  in  a more evolved
massive star forming  region.  Thus because the  massive  protostellar
candidates are  the  first massive stars  to  form within   their host
molecular clouds, any associated clusters should represent clusters in
their earliest evolutionary phase.

Our study is similar in its objectives to a previous survey of stellar
clusters       around     Herbig     Ae/Be     stars      (Testi    et
al.~\cite{testi97},~\cite{testi99}).    That study   also   sought  to
identify very   young clusters with  active   star formation using the
Herbig Ae/Be stars, which are protostars, as indicators of active star
formation.  In our study we use the massive protostellar candidates as
indicators  of the  earliest stages  of  stellar and therefore cluster
formation.

A number  of  independent  near-infrared  (NIR)  observations  of  the
regions around several  of the massive  protostellar candidates in our
target  list   already show evidence   for  embedded  stellar clusters
(references in Table 2).  In contrast to these previous  observations,
in this study, we undertake a  systematic search for embedded clusters
around all the previously identified high mass protostellar candidates
in  the lists of   Molinari  et al.~(\cite{mol96})  and  Sridharan  et
al.~(\cite{sri01}).

We identify the potential clusters  as star count density enhancements
above  the mean background  level (Lada \& Lada~\cite{ll95}; Carpenter
et al.~\cite{car00}; Ivanov et al.~\cite{ivanov02}) using the existing
K-band      observations  of   the   2MASS    database  (Kleinmann  et
al.~\cite{2mass94}).  We  report on the  identification of 54 clusters
by this technique of which  34  are new  detections. We estimate  some
basic  properties   that  can be   derived  from  the   J,H,K  data in
2MASS. Finally we discuss some implications of  newer theories of high
mass star formation for the formation and evolution of clusters.

\section{Method} 

The 217 targets selected for our study represent the unique sources
from the combined lists of 163 candidates in Molinari et
al.~(\cite{mol96}) and 69 in Sridharan et al.~(\cite{sri01}). We
searched for clusters within an area of $400\arcsec{} \times
400$\arcsec{} around each of the 217 positions of high mass
protostellar candidates.

The 2MASS database contains J,H, and K bands, and we used the K-band
at 2.2$\mu$m for the initial identification of the clusters because
K-band suffers the least extinction.  We constructed stellar surface
density maps by spatially binning the 2MASS point sources. We selected
sources with quality flags of A,B,C, and D, excluding lower quality
sources.  We used bins of 120\arcsec separated by 60\arcsec to cover
the 400\arcsec$\times$400\arcsec area.  Clusters were detected as star
count density enhancements above the mean background level within the
sampled area.  We used the average noise($\sigma$) in the map as a
reference and plotted contours of mode$+$2$\sigma$ and above to
identify clusters.  The mode value represents the average star count
in the region, and mode$+$2$\sigma$ reveals the enhancement with
respect to the average background.  We identified 63 possible clusters
for which we retrieved data over a larger area of
600\arcsec$\times$600\arcsec~ in all the three photometric bands to
enable better sampling of the background counts and also to construct
color-color diagrams and color-magnitude diagrams. Based on improved
sampling of the background that reduced the noise in the contour maps,
and iterating with different bin sizes, we excluded some targets that
had loose groupings of 5-8 stellar sources. After excluding
such weak groupings, we identified 54 clusters all of which contain
more than 8 stars, although the majority of the clusters show star
counts in excess of 20-30. We then used a smaller bin-size of
80\arcsec (step size 40\arcsec) to reveal relatively small scale
structures within the detected clusters. The resulting contour maps
for each of these 54 clusters are presented in Figure.~1.  The contour
levels begin at mode+2$\sigma$ and increase in steps of 1$\sigma$.

In order to estimate the number of stars in each cluster we must
account for the inclusion of background and foreground stars within
the defined cluster boundary. To estimate the combined background and
foreground stars, we sampled for each cluster an adjacent region of
1600$\arcsec^{2}$-2500 $\arcsec^{2}$.  The corrected counts are listed
in Table.~2 as ``true'' cluster membership. If we assume that within
the sensitivity of 2MASS data we might not detect any stars behind the
high extinction of a massive star forming cluster, then our
subtraction of the estimated count of background stars would result in
an underestimate of the number of true cluster members. Thus our
estimate of the numbers of stars in a cluster might represent a lower
limit to the true number in cases where the extinction through the
clusters is very high.

\subsection{Detection statistics}

We detected 54 clusters out of 217 unique sources which implies an
overall $\sim$25\% detection rate.  However, there are no clusters
detected in the RA range 6hr to 20hr that coincides with the Galactic
mid-plane although some targets were associated with K-band
nebulosity.  Our technique of identifying clusters as stellar density
enhancements in the 2MASS data does not appear suitable to detect
clusters in the Galactic mid-plane.  First, the high extinction in the
Galactic mid-plane reduces the number of stars that we can detect.
Second, in the Galactic mid-plane the background count level is high
and star count density enhancements are less significant with respect
to the background. Excluding the targets in the 6 -- 20 hr RA range,
the detection rate for clusters around the massive protostellar
candidates is 60\%.

\begin{figure*}
\centering
\includegraphics[width=\textwidth]{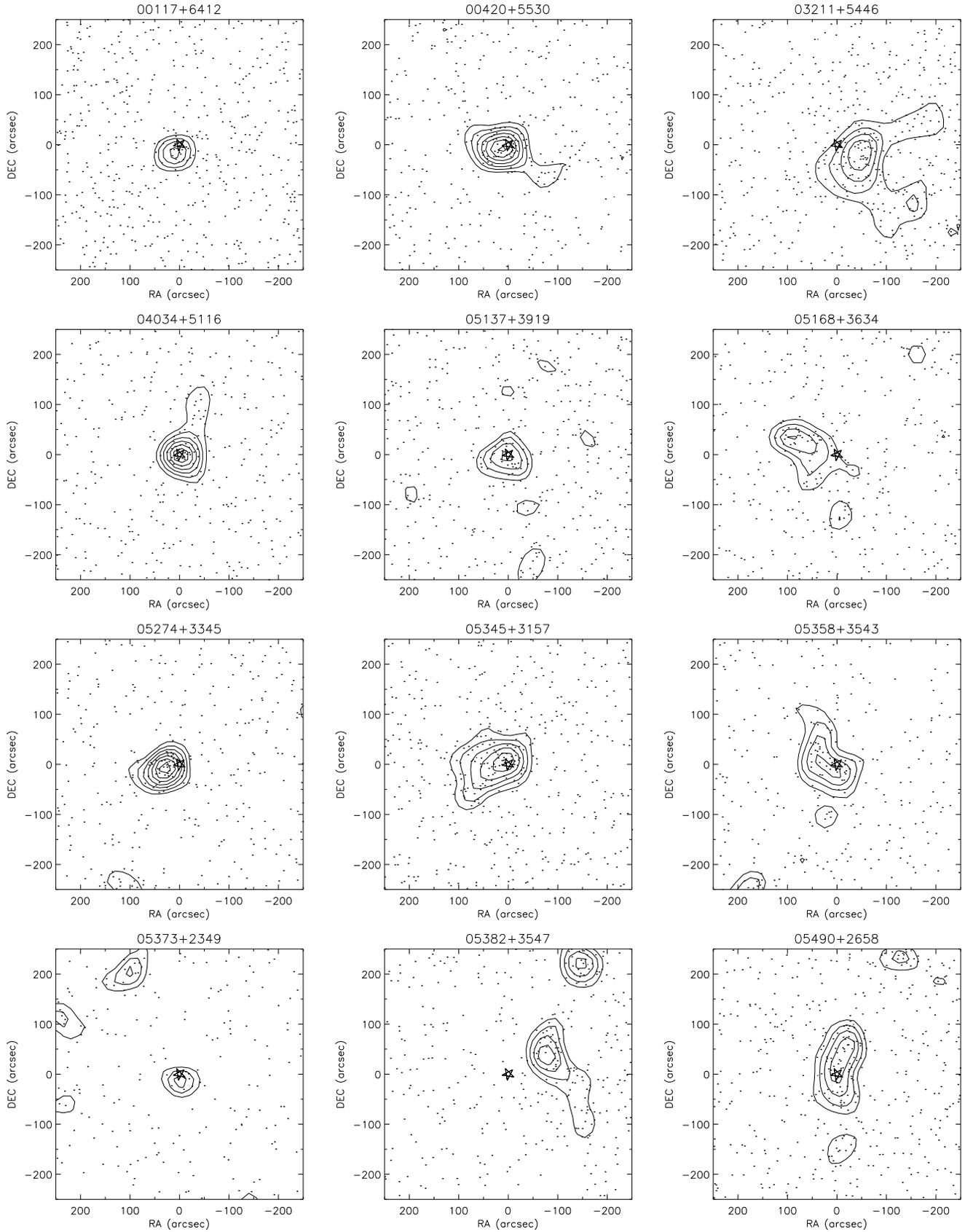}

\caption{Stellar surface density contour plots for the 
54 embedded clusters produced with a bin size of 80$\arcsec$(step size
= 40\arcsec). The star symbol represents the position of the luminous
IRAS source. Contours begin at mode+2$\sigma$ level and increase in
steps of 1$\sigma$, $\sigma$ being the average noise measured over the
an area of 600\arcsec$\times$600\arcsec centered on each target. The
dots represent the point sources with phqual=A,B,C,D catalogued in
the 2MASS GATOR database. }

\label{figure1}
\end{figure*}
\addtocounter{figure}{-1}
\begin{figure*}
\centering
\includegraphics[width=\textwidth]{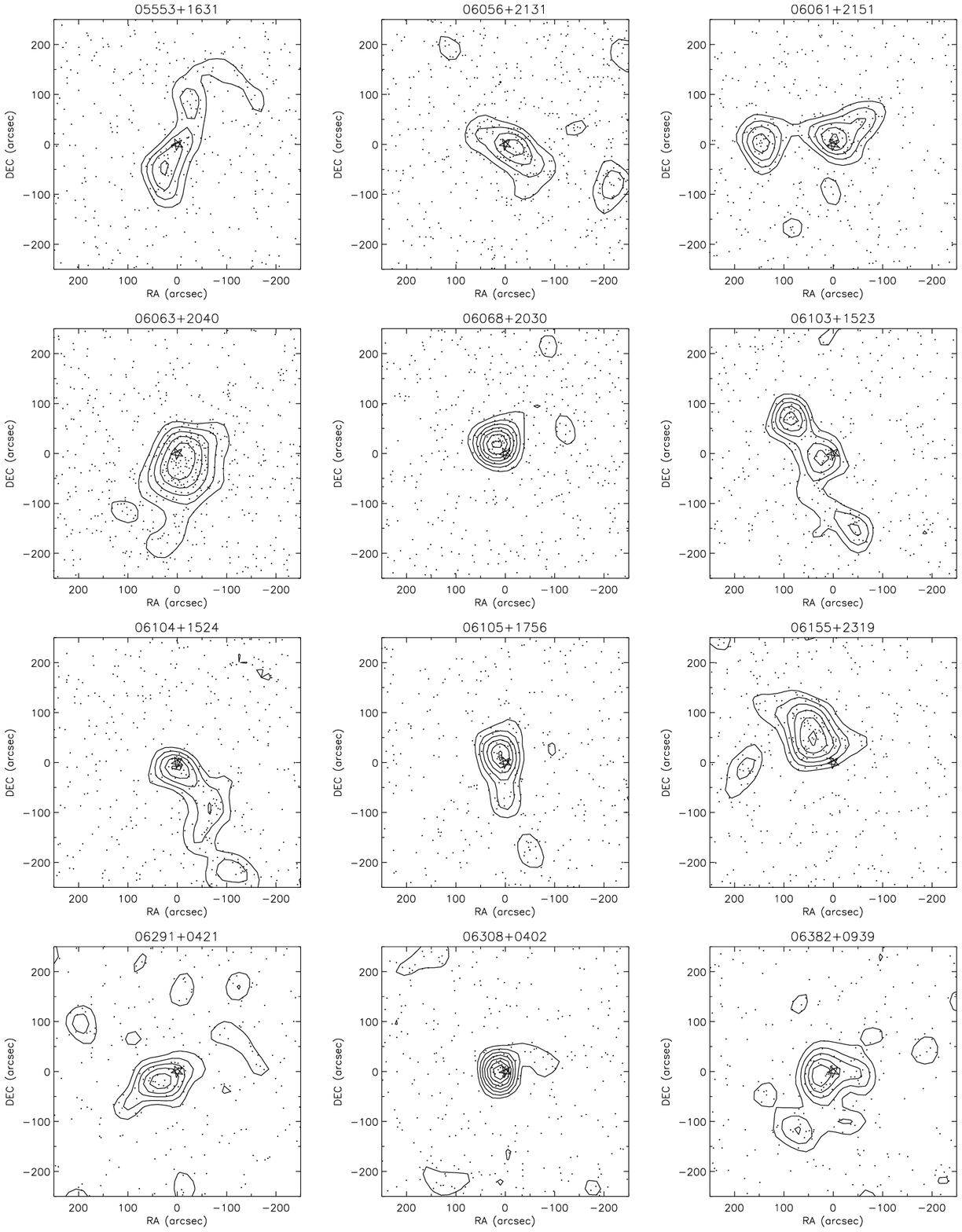}
\caption{contd. Stellar surface density contours of detected clusters} 
\end{figure*}
\addtocounter{figure}{-1}
\begin{figure*}
\centering
\includegraphics[width=\textwidth]{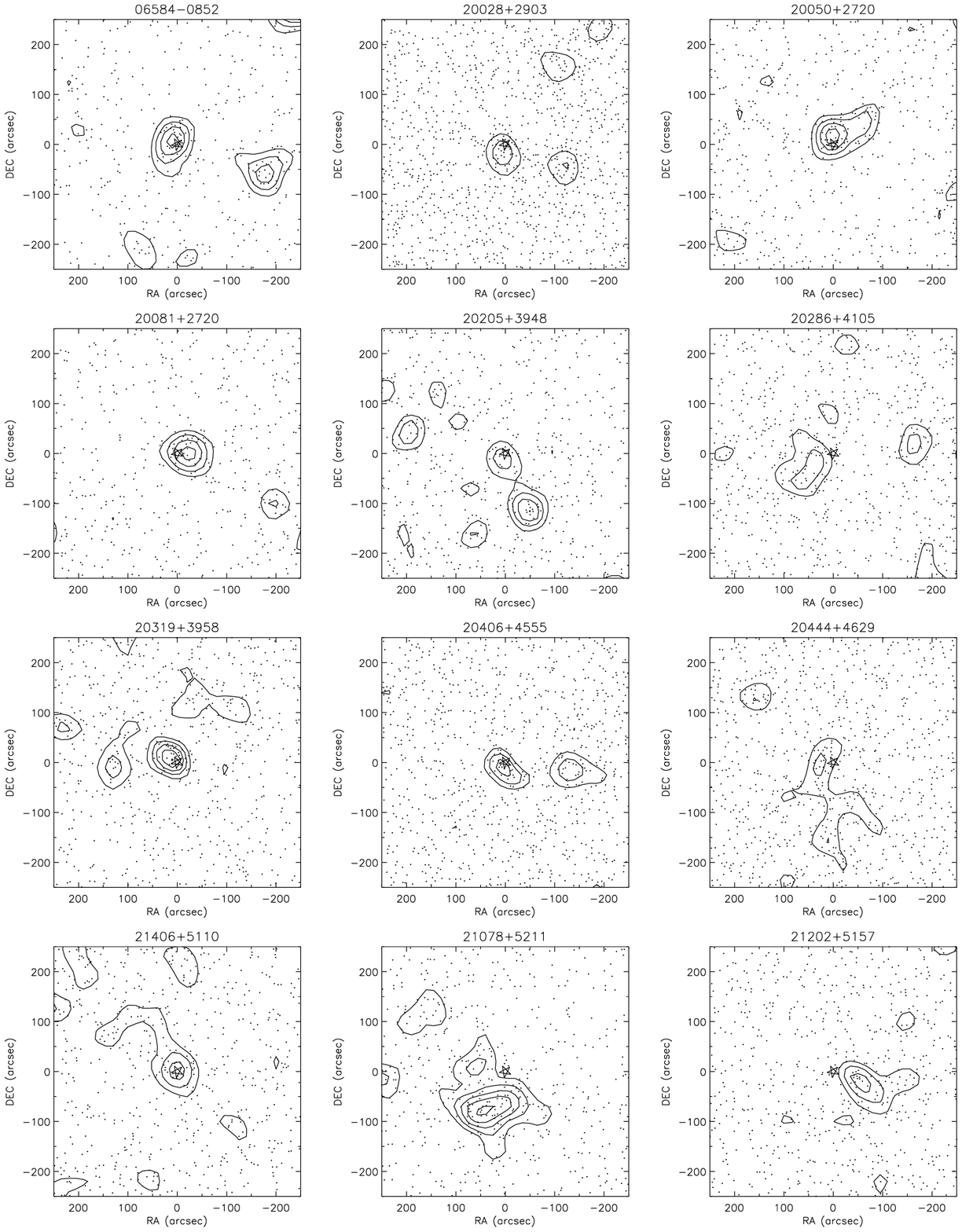}
\caption{contd. Stellar surface density contours of detected clusters}
\end{figure*}
\addtocounter{figure}{-1}
\begin{figure*}
\centering
\includegraphics[width=\textwidth]{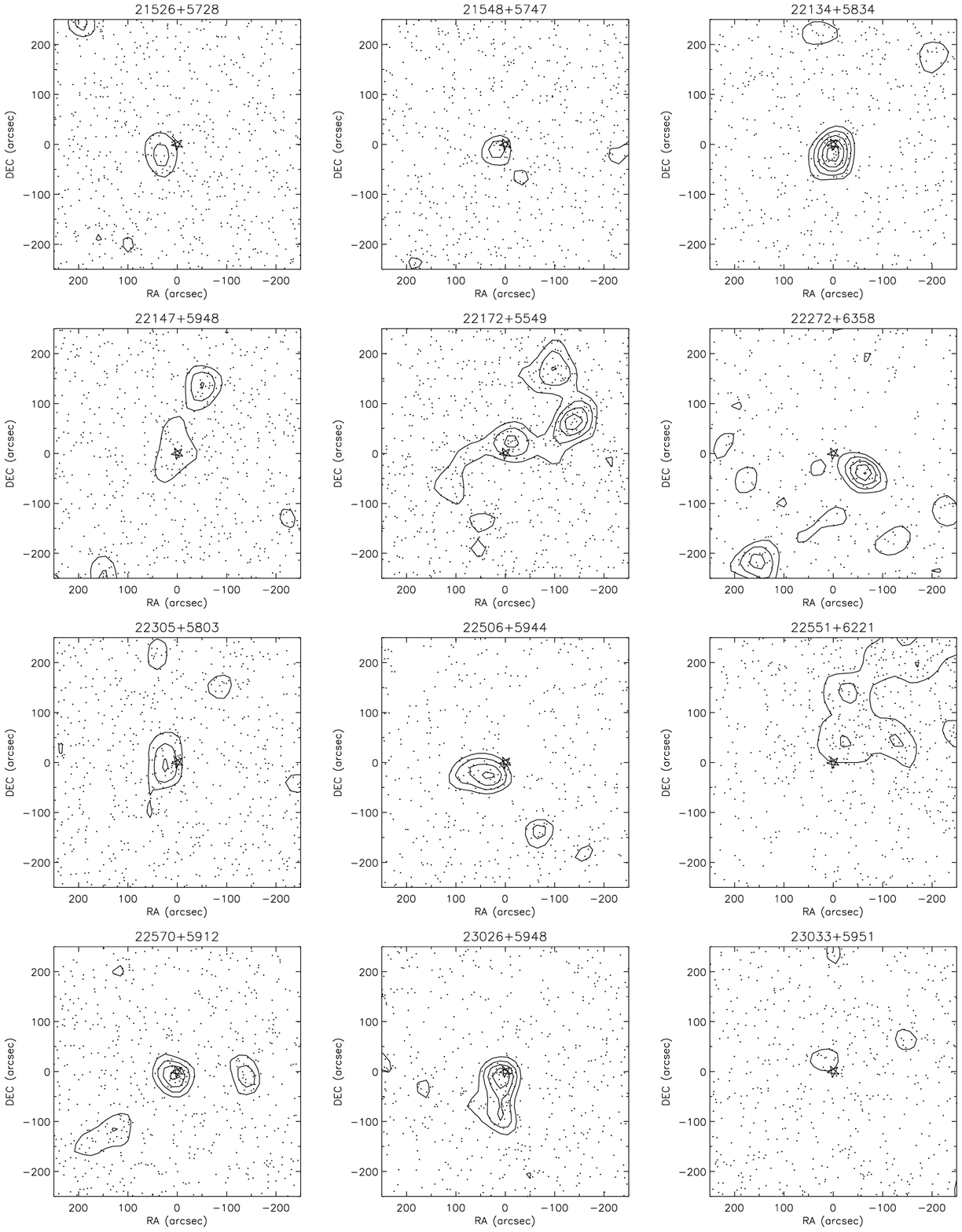}
\caption{contd. Stellar surface density contours of detected clusters}
\end{figure*}
\addtocounter{figure}{-1}
\begin{figure*}
\centering
\includegraphics[width=\textwidth]{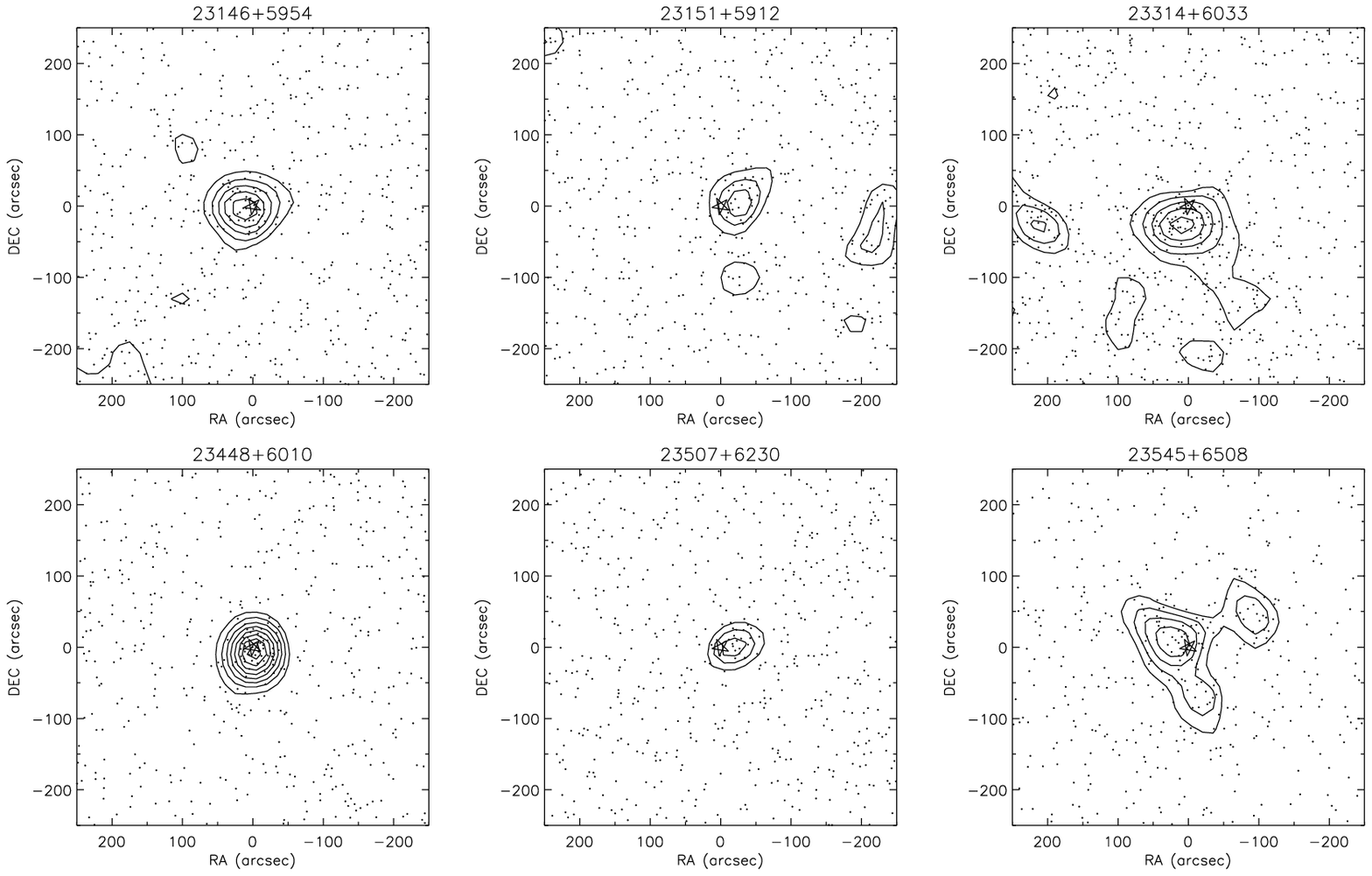}
\caption{contd. Stellar surface density contours of detected clusters}
\end{figure*}

\subsection{Cluster membership and properties}

In Table.\,1 we list the radius, distance, and luminosity of each
cluster.  The radius is defined from the area enclosed by the
mode$+$2$\sigma$ contours as ${\rm R} = \sqrt{area}$/$\pi$.  The
distance and luminosity are generally quoted from their source papers
namely MBCP96 and Sri01. Where available, improved distance estimates
are used based on a SIMBAD search. In particular, targets beyond the
Solar circle are relatively less studied making their distance
estimates limited to the values quoted by Wouterloot et
al.~(\cite{wbf93}). A histogram of the effective radii of the clusters
indicates a mean cluster radius of $\sim$1\,pc. The IRAS peak
positions and the cluster peak positions coincide in about half of the
detected sample and do not coincide in the remaining half. However,
the fraction in which the two peaks coincide, the clusters are noted
to be densest and are more circularly defined than the rest.

\subsection{Cluster mass estimation}

We estimated the mass of each cluster based on the method described by
Lada \& Lada (\cite{ll03}, Hereafter LL03). The method incorporates
the K-luminosity function (KLF) models of Muench et al.
(\cite{muench02}) and the evolutionary tracks of D\'Antona \&
Mazitelli (\cite{dm94}).  The mass estimation is done by posing the
following question. If we were to place the Trapezium cluster at the
same distance as one of the sample clusters and subject it to the same
dust extinction, how many of its 780 members (see Muench et
al. \cite{muench02}) are we likely to observe?  This question can be
trivially answered if we know the completeness limit (limiting
magnitude) of our observed data and by using D\'Antona \& Mazitelli
(\cite{dm94}) tracks along with the formula\\ M$_k$ = K$_{mag}$ - 5 +
5 log(distance) - A$_k$. \\ We then compute a ratio of the observed
number of stars to the expected number if our cluster were the
Trapezium personified. This ratio then merely indicates the ratio of
the mass of the sample cluster to that of the mass of the Trapezium
cluster which is assumed to be 413\,M$_{\odot}$ (Muench et
al. \cite{muench02}).

We use the individual extinction values to each cluster as derived
from the color-color diagrams (see Sec\,2.5) of each cluster. Further,
the 2MASS limiting magnitude in K-band of 15mag and an average cluster
age of 1Myr is adopted to compute the masses.  The resulting mass
estimates are listed in Table.\,1.  Fig.\,2 shows the embedded
cluster mass distribution function (ECMDF) constructed using the
cluster mass estimates in Table.\,1.  The solid line represents the
ECMDF for the sample from our work and the dotted line shows the LL03
ECMDF for their sample of embedded clusters within 2 kpc from the
Sun. The ECMDF of both samples are similar.

In Fig.\,.3 we  plot the IRAS luminosity  of the targets  versus the
mass of    the associated embedded  cluster.   There  is a correlation
between the mass of the associated embedded cluster and the luminosity
of the  target massive protostar, similar to  the correlation found in
HAeBe stars by Testi et al.~(\cite{testi99}).

\subsection{The morphology of embedded clusters}

In their review of embedded clusters, Lada \& Lada (2003) suggest that
the embedded clusters can be classified into two morphological types,
hierarchical and centrally condensed, and they suggest that these
structures may reflect the physical processes responsible for cluster
formation.  In particular, the two types may indicate a dominance of
turbulent (hierarchical) or gravitational (centrally condensed)
energies.  The embedded clusters, and particularly the youngest
clusters still in formation, offer the opportunity to test this
hypothesis against observations. Such a test would best be done
comparing the structure of the stellar distribution of the clusters
with the gas distribution in the clouds. This comparison would require
observations outside the scope of this current research.  However, we
may compare the relative numbers of hierarchical and centrally
condensed clusters in the two samples of embedded clusters, those
identified in our current research as the youngest by their
association with the massive protostellar candidates and those
embedded clusters within 2 kpc of the Sun, from the list in Lada \&
Lada (2003).  We classify the clusters that show single peaks in the
stellar density maps as centrally symmetric (C-type) and those which
show more than one peak as hierarchically structured (H-type).  The
results for our sample of clusters (Fig.~1) indicate a ratio
H-type/C-type $\sim$0.8. We applied the same classification scheme to
the 70 embedded clusters of Lada \& Lada (2003), using stellar density
maps produced from the 2MASS data by our procedures as described in \S
2. The nearby embedded clusters have a ratio H-type/C-type $\sim0.9$,
essentially the same as their younger counterparts.

\begin{figure} 
\resizebox{\hsize}{!}{\includegraphics{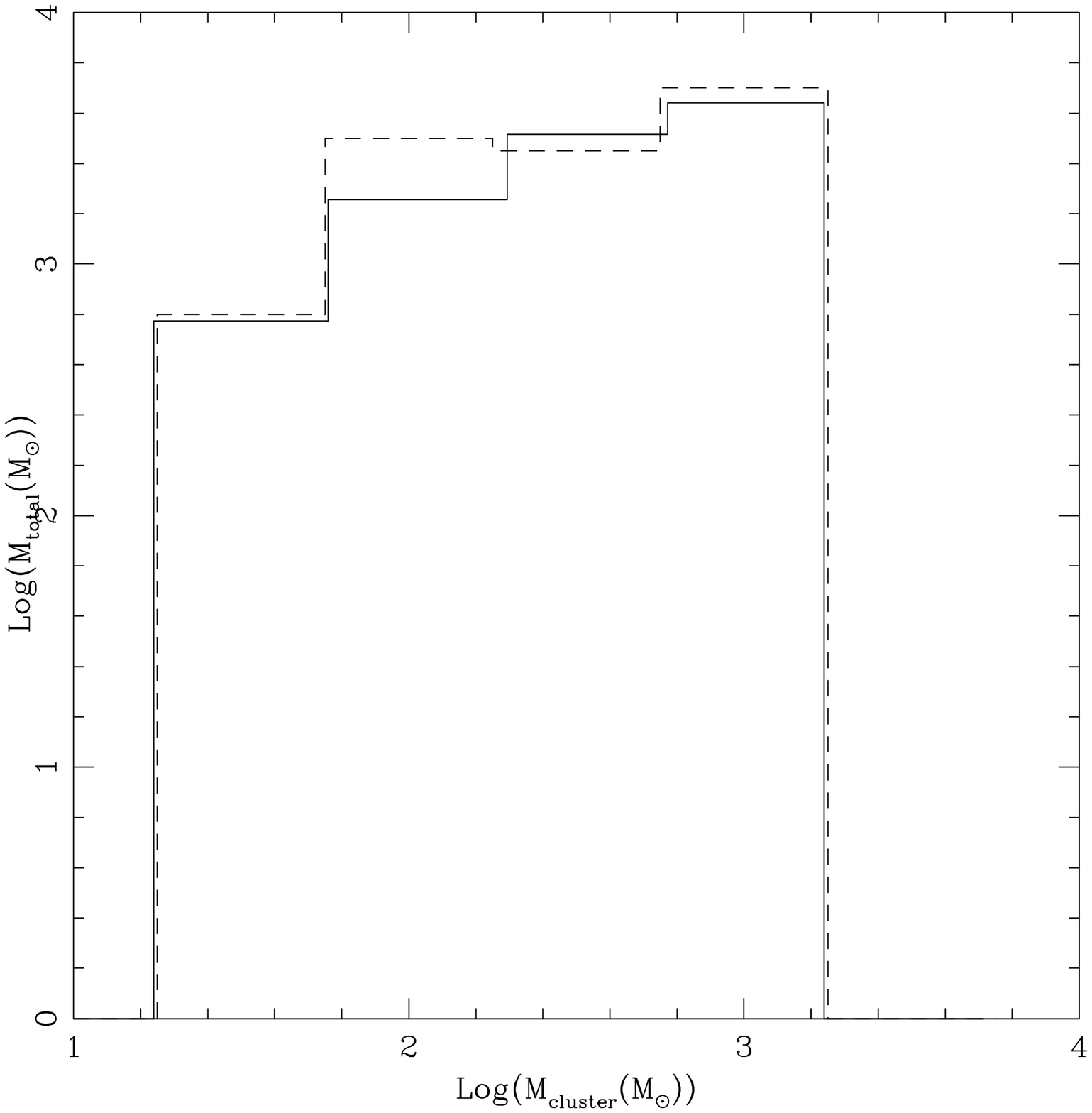}}
 \caption{Embedded cluster mass distribution function. The dotted line
 represent the ECMDF from LL03 for the embedded clusters within 2\,kpc
 distance. The solid line shows the ECMDF for the clusters from this work.}

 \label{fig:2} 
\end{figure}

\begin{figure} 
\resizebox{\hsize}{!}{\includegraphics{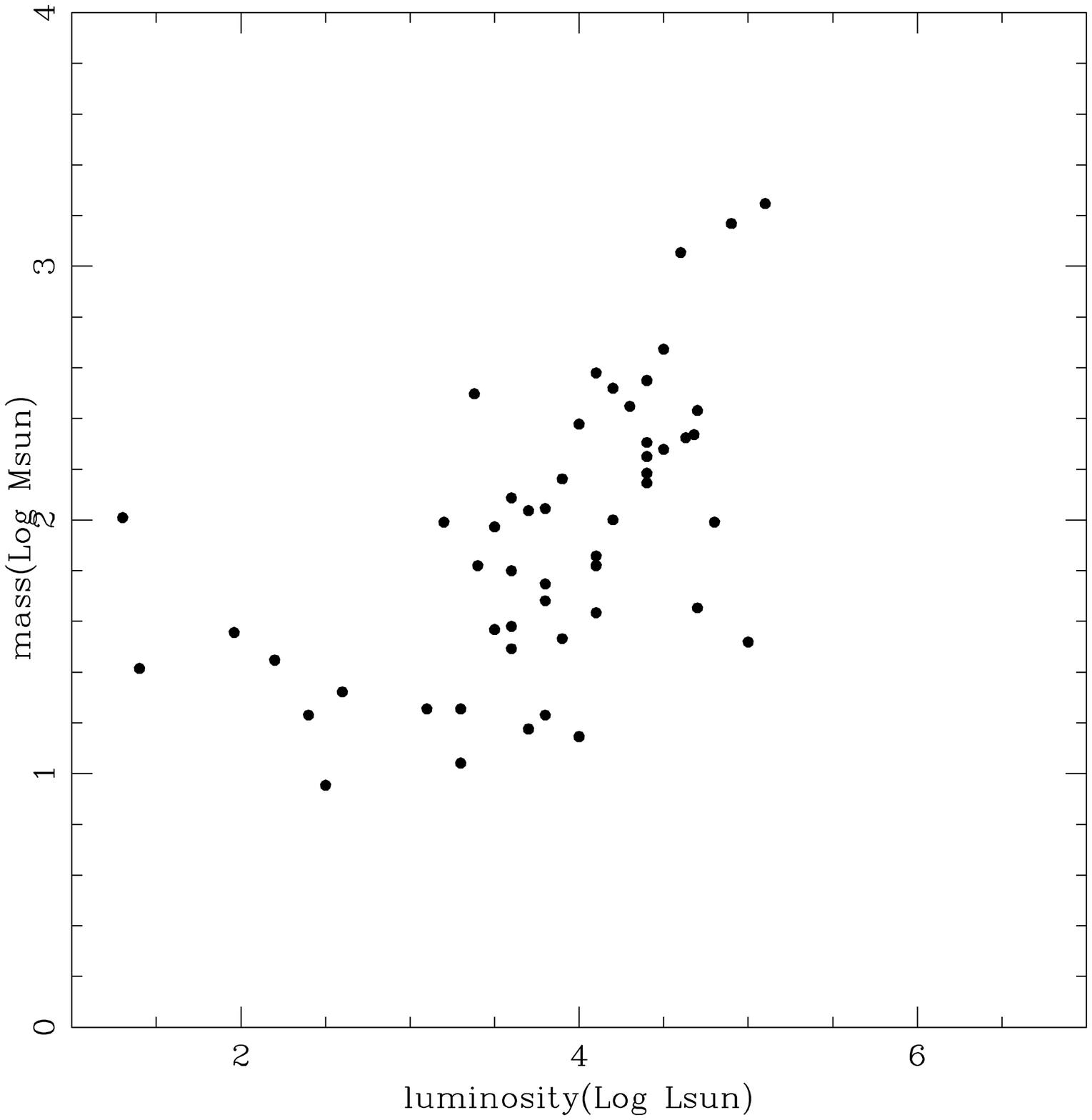}}
 \caption{A plot of the mass of the associated clusters vs the
 luminosity of the target massive protostar}

 \label{fig:3} 
\end{figure}

\subsection{Photometric analysis}

Once the clusters were detected, we constructed H-K vs J-H color-color
and H-K vs K color-magnitude (CM) diagrams for each of the clusters to
investigate the nature of the associated point sources. To do this, we
first transformed the 2MASS data into the Bessel \& Brett system using
the transformations: 
\\ (H-K)$_{BB}$ = 1.0298 $\times$
(H-K)$_{2MASS}$-0.0350 \\ (J-H)$_{BB}$ = 1.0101 $\times$
(J-H)$_{2MASS}$-0.0495 \& \\ K$_{BB}$ = 1.0298 $\times$
K$_{2MASS}$-0.0350.\\
We plotted all points
that fell within the mode+1$\sigma$ contour to allow consideration of
even those members that fell at the extreme edges of the clusters along with
the main-sequence dwarf and giant tracks, reddening vectors, T-Tauri
locus (Meyer et al. \cite{meyer97}), and the HAeBe locus (Lada \&
Adams \cite{la92}).  We dereddened the
points that fell within the reddening vectors enclosed by the
main-sequence tracks to the M6-K6 locus in order to find the
extinction to each point.  A histogram of all such extinction values
and an average extinction was constructed for each cluster.  These
values are listed in Table.~2 and are used to obtain the
corresponding mass estimates for each cluster.  We also constructed
H-K vs K color-magnitude diagrams for each cluster and verified the
consistency of extinction derived from color-color diagrams.

\begin{figure} 
\resizebox{\hsize}{!}{\includegraphics{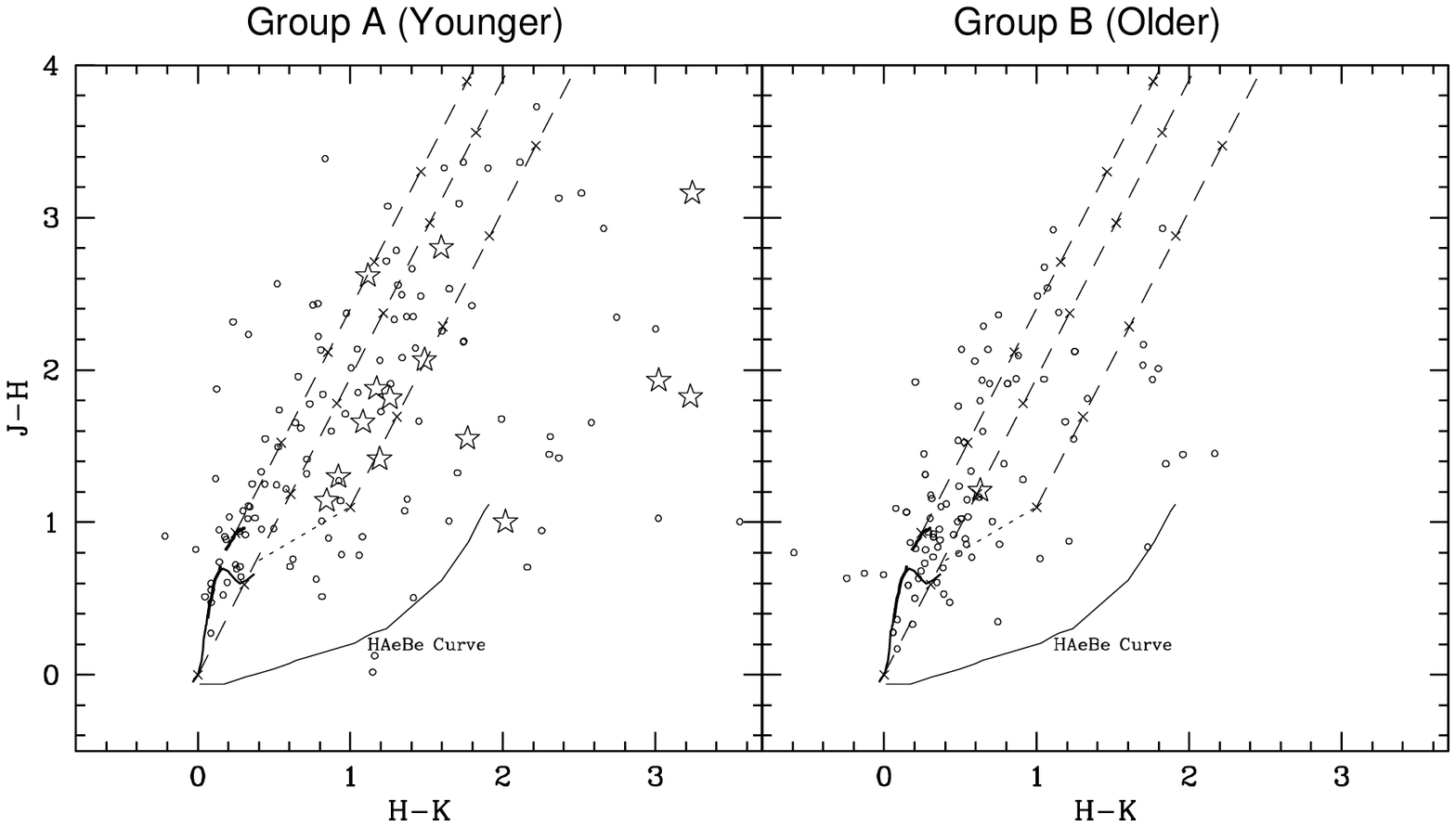}}

 \caption{ Color-color diagram of NIR counterparts to the 1.2\,mm
 peaks of HMPOs. The solid curve represents the HAeBe curve by Lada \&
 Adams (\cite{la92}), thick-solid curve represents the main-sequence
 tracks, the dotted line shows the T-Tauri locus from Meyer et
 al.(\cite{meyer97}) and the dashed lines represent the reddening
 vectors. The dots represent the point sources from GATOR PSC and star
 symbols represent the extended sources from the GATOR XSC.}

 \label{fig:4} 
\end{figure}

\subsection{The NIR colors of massive protostars}

The massive protostellar candidates were all initially identified by
their IRAS colors, and the list further refined by the association of
the candidates with other signs of star formation (references in
Introduction). We may ask whether the NIR colors of any of the
clusters or their members are distinct.

For example, in the formation of high mass stars, it has been
hypothesized that the dense accretion flows that form the stars
suppresses the formation and later the expansion of an HII region
(Walmsley 1995, Keto 2003). Once the HII region reaches a size
detectable by typical VLA observations, the accretion flow will have
ended and the star should have obtained its main sequence mass. The
list of massive protostellar candidates contains candidates associated
with radio continuum emission (HII regions) and those without
detectable radio continuum. In order to identify any differences
between the NIR emission of these two types of sources, we searched
for NIR emission within 5" of the dust continuum peaks mapped by
Molinari et al.~(2000) and Beuther et al.~(2002) using both the 2MASS
Point Source Catalog (PSC) and Extended Source Catalog (XSC).  The
extended NIR sources are shown as star symbols, and NIR point sources
are shown as points in Fig.\,4 plotted along with the main sequence
dwarf \& giant loci, the T-Tauri locus and the HAeBe locus.  Reddening
vectors assume an interstellar reddening law of R=3.12.  The sources
with no radio continuum (Group A in Fig.~4) display very red colors,
are spread around the region of the main sequence and T-Tauri loci,
and extend way beyond the limits of the HAeBe locus.  In comparison,
the older sources are limited to the main-sequence region with some
falling within the HAeBe locus. As far as this test goes, the data are
consistent with the theory that the massive protostellar candidates
without associated radio continuum are more deeply embedded than those
with HII regions.  The test is limited. The NIR emission, particularly
the extended emission, is not necessarily associated with any massive
protostellar candidate. The NIR emission is from regions of very high
extinction that may affect the J,H, and K magnitudes. Spectroscopic
observations would be more reliable in identifying individual
sources. However, the J,H, and K data available in the 2MASS survey in
combination with the continuum emission from dust may be sufficient to
suggest the locations of the massive protostellar candidates without
suggesting a correspondence with individual stars.

\section{Discussion}

\subsection{Implications for the  relative ages of low and high mass stars}

If one   accepts   the hypothesis  that   the  high mass  protostellar
candidates  identified by the  IRAS satellite can  be localized by the
observed combination  of  the dust  continuum and  NIR, then there  is
evidence for  mass  segregation in  our sample of  the youngest clusters
still in formation.  Furthermore if  we accept the hypothesis that the
massive protostellar candidates are the first massive stars to form in
a  cluster,  the presence of the   cluster itself, which is identified
primarily through its more numerous lower  mass members, suggests that
the low mass  stars  form before  their high mass  counterparts.  This
hypothesis has   been in    the  literature   for  quite    some  time
(e.g.~Herbig~\cite{herbig62}).  The clusters   in our study,  by their
association with massive   protostars, reiterate  the suggestion  with
slightly different reasoning  than presented earlier (e.g.~Hillenbrand
et al.~\cite{hil93}; Testi et  al.~\cite{testi99}).  The difference is
that,  in  our sample  the sources are  relatively  more  massive than
typical  HAeBe stars, $\sim$40\% of   the  detected clusters are  also
associated with massive outflows (see  Table.~2) and many sources lack
any significant HII regions. Therefore, the massive stars are still in
accreting phases  as evidenced by   their outflows while  the low mass
members of the associated clusters have finished accreting and emerged
into the  Class I    or II  phases  as  evidenced  by  their   2$\mu$m
appearance. Therefore our result places a stronger constraint that the
massive stars begin to form at least after  a few 10$^4$ yrs after the
first low mass stars in these embedded clusters were born.

\subsection{Implications for the formation mechanism of massive stars}

The  previous study of stellar clusters   associated with Herbig Ae/Be
stars by  Testi et al.~(\cite{testi97}, \cite{testi99}) suggested that
the stellar spectral type of  the Herbig Ae/Be  star may be correlated
with the richness (stellar density) of the  cluster. This leads to the
hypothesis that dynamical   interaction (collisions) between  low mass
stars  may  be  required to    form  a  high   mass star   (Testi   et
al.~\cite{testi99}). That hypothesis was motivated in part by the high
stellar densities observed.   Our   study  does not contradict    this
suggestion, but our results also suggest an alternative hypothesis.

In our sample, those clusters in which the IRAS/mm peaks and the
cluster peaks coincide most closely are also the densest clusters both
in terms of stellar density as well as molecular gas density.  The
very high gas densities in the centers of these clusters, inferred
from observations of maser and thermal molecular line emission and
radio continuum emission are consistent with the hypothesis that the
high mass stars form by continuing accretion onto lower mass stars.
The hypothesis of continuing accretion requires very massive accretion
flows, perhaps up to $10^{-3}$ {\rm M}$_\odot$ yr$^{-1}$ to form the
earliest type stars.  (Keto~\cite{keto03}). These massive flows are
more likely found in very dense gas that our study, in conjunction
with previous observations, locates at the centers of the clusters
where the most massive stars are forming.

There are three principle reasons why the the hypothesis of continuing
accretion requires a very high rate of accretion to form an early type
star. First, the rate of accretion must be high enough  to keep a star
from evolving off the main sequence before it  has reached a high mass
(Keto \cite{keto03}).  Second,  the   momentum of the  accretion  flow
$\dot M v$ must be high enough to overcome the high radiation pressure
of the stellar  luminosity (Kahn \cite{kahn74};  Wolfire \& Cassinelli
\cite{wolcas87}; Keto  \& Wood \cite{ketowood05}).  Third, the density
of the gas around the star must be high enough to  keep the HII region
around  the star small  so that the accretion flow  is not reversed by
the thermal  pressure of the ionized  gas.  Specifically the radius of
ionization equilibrium  must be within the   distance where the escape
velocity from   the  star equals    the   ionized sound  speed   (Keto
\cite{keto02a},\cite{keto02b}).

The hypothesis of continuing accretion has specific implications for
our understanding of stellar evolution and therefore cluster
evolution. It is a fundamental assumption of the standard theory of
stellar evolution that once the mass of a star is fixed, its
subsequent evolution is determined. Stars that continue to gain mass
by accretion obviously do not have a fixed mass, and their
evolutionary tracks in color-magnitude space or the Hertzprung-Russel
diagram are different than for non-accreting stars, and furthermore
depend on the rate of accretion.  For example, the stellar structure
calculations of Stahler, Shu \& Taam
(\cite{Stahler1980a},\cite{Stahler1980b},
\cite{Stahler1981}) showed that stars of   mass greater than about   7
M$_\odot$ begin  hydrogen burning while still  in  the accretion phase
and thus do not have a protostellar  phase equivalent to that in lower
mass stars. This result has been  confirmed in subsequent calculations
(Beech and   Mitalas~\cite{bm94};    Behrend  \&   Maeder~\cite{bm01};
Bernasconi  and    Maeder~\cite{bm96};            Meynet           and
Maeder~\cite{mm00}).  Second, continuing  accretion will supply  fresh
hydrogen  to the star,  and  if the  rate is   high enough, the  fresh
hydrogen will prevent  the star from evolving  off  the main sequence.
Third for  stars below about  7 M$_\odot$, the accretion  will provide
fresh deuterium to  the star and prolong the  phase of deuterium shell
burning,  altering  the  pre-main  sequence evolutionary  tracks.   We
briefly discuss these   differences in  stellar   evolution and  their
consequences for our understanding of cluster evolution.

The accretion rates and time scales necessary for an intermediate mass
star to evolve to a high mass star can be estimated through stellar
structure calculations. Figure.\,5 shows evolutionary tracks for stars
with continuous accretion at various rates.  The evolutionary tracks
were calculated for us by Alessandro Chieffi based on the stellar
structure models described in Chieffi, Staniero \& Salaris
(\cite{chief95}). The evolutionary tracks are calculated for accretion
rates normalized to $6\times 10^{-8}$ M$_\odot$ yr$^{-1}$ for a star
of unity M$_\odot$ but with different dependencies on the stellar mass
in each track. While the mode of accretion in high mass star formation
is not well known, observations indicate that the rates may scale as a
power of the stellar mass (Churchwell \cite{church98}, Henning
\cite{hen00}, Behrend \& Maeder \cite{bm01}). We plot accretion rates
scaling with powers of the stellar mass from 0.5 to 2.0. A rate
proportional to the square of the mass is the maximum accretion rate
attainable in purely spherical accretion (Bondi \cite{bondi52}).

\begin{figure} 
\resizebox{\hsize}{!}{\includegraphics[angle=90]{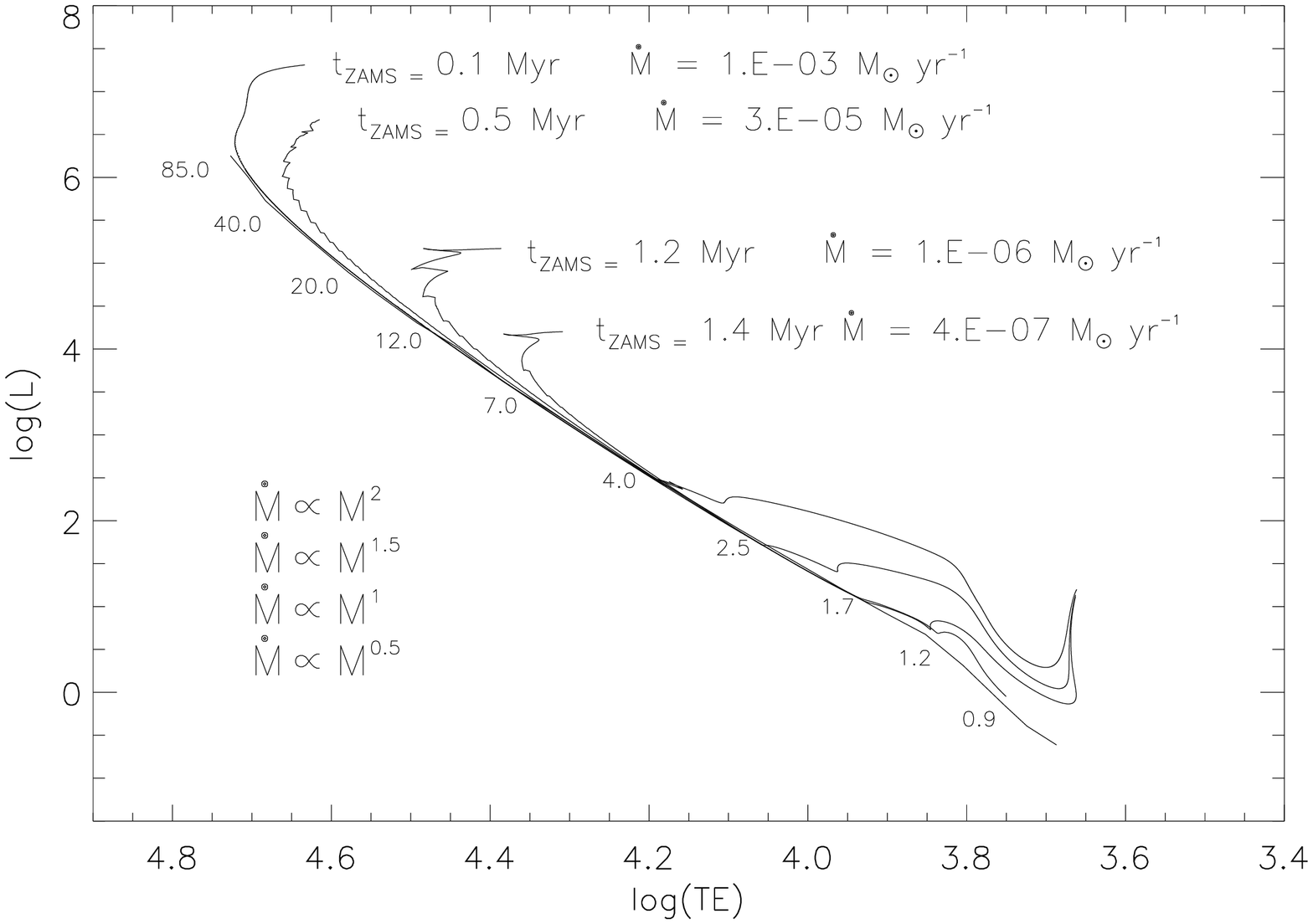}}

\caption{An HR-diagram showing evolutionary tracks for massive stars 
that are gaining mass by accretion. The zero age main sequence (ZAMS)
line is shown annotated by stellar mass in solar masses. The four
evolutionary tracks are for stars accreting with rates ($\dot{M}$)that
vary as a power law of stellar mass (M), with indexes 0.5,1,1.5 \& 2
(bottom to top respectively).$t_{\rm ZAMS}$ indicates the time spent
on the ZAMS line for a 4~M$_{\odot}$ star before it deviates off the
line. The accretion rate at the time of deviation, which is also the
maximum accretion rate for that track is marked below each track.}

 \label{fig:5} 
\end{figure}

The  time scale for   the  formation  of  high  mass stars  is   quite
short. The  time required  for a  star  of 4.0 M$_\odot$ to  reach the
point where  it turns off the main  sequence toward the giant branch is
printed on  the  figure for  each accretion  rate.   The main sequence
lifetimes for  stars of B to  late O type  are roughly between 0.5 and
1.0  Myr.  The most massive  O  stars spend such  a  short time in the
accretion phase  that  observed  examples  of early   O  stars  in the
accretion  phase would  be expected   to be  quite rare.  Because  the
accretion rate scales as a power of the stellar mass, these precursors
to massive stars   spend most of their time   in the lower mass  range
where the   accretion rate is  lower and  the  evolution proceeds more
slowly. Thus we would  not expect to see  many O type stars  in deeply
embedded clusters that are at the earliest stages of formation.

Continuing accretion  supplies fresh deuterium   to the stars allowing
deuterium burning in the outer shell of the star  to continue past the
time when  the deuterium would have  been exhausted in a non-accreting
star.  This keeps  the stellar radius   larger and thus the luminosity
higher  than   for  non-accreting   stars.   Thus    pre-main sequence
evolutionary tracks of accreting stars maintain a higher luminosity as
the pre-main sequence star evolves (Fig.~5; see also Behrend \& Maeder
\cite{bm01}).  At  around   4~\msun,  the luminosities    of both
accreting and non-accreting stars are no longer dominated by Deuterium
shell burning and  their luminosities are  essentially the same.  Thus
as accreting stars above 4~\msun gain mass they appear  to move up the
main  sequence  as  defined by  the  standard  theory of non-accreting
stars. If all massive stars in  a cluster form by continuing accretion
onto lower mass stars, then there will be no massive pre-main sequence
stars and  no stars   above 4~\msun in  the  HR  diagram on high  mass
pre-main sequence tracks.   HR diagrams from  observations of clusters
are consistent with this hypothesis (NGC 2264,  NGC 6530 Stahler 1985;
NGC 6611 Hillenbrand et  al.~1993). In particular, these clusters show
large numbers of stars with  masses less than   4 M$_\odot$ above  the
main sequence, but few stars above the main sequence at higher masses.
These HR diagrams   are of course  also  consistent with the   stellar
evolution of  non-accreting   stars.  In particular,  the   absence of
observed pre-main  sequence massive stars could  be due to the brevity
of  the pre-main sequence  phase in  massive stars.  Nevertheless, the
determination of the ages of stars and clusters by the location of the
stars on the HR diagram already has some inherent uncertainty (Stahler
1985),   and the possibility   of  continuing  accretion adds  further
uncertainty to this method of age determination. Thus a method to date
clusters  by  their  association  with  protostars   as in  Testi   et
al.~(\cite{testi99}) or by their association with massive protostellar
candidates as in this study should be useful in distinguishing between
the different   hypotheses  of  accreting   and non-accreting  stellar
evolution.

\subsection{Observational evidences for continuing accretion}

Finally, we note that several individual studies of massive protostars
in recent years support the conjecture that massive stars are most
likely formed by continuing accretion. First, continuing accretion
requires that an intermediate mass star such as an A or B star forms
by accretion and should be surrounded by an envelope mass much greater
than the protostar itself, so that sufficient material is available to
feed the intermediate mass star to build its mass.  Second, the
accretion rates should be high enough of the order of 10$^{-3}$~\msun
yr$^{-1}$.  Detailed studies of sources such as IRAS20126+4104
(Cesaroni et al. \cite{cesa94}), IRAS20293+3952 (Beuther et
al.\cite{beu04}), IRAS07427-2400 (Kumar et al.
\cite{kumar03b}),  G31.41+0.31  and   G24.78+0.08   (Beltran  et   al.
\cite{beltran05}) have shown the presence  of both Keplerian disks and
large toroids that contain several  hundreds to  a few thousand  solar
masses of  material.  In the case  of IRAS07427-2400, the massive star
is visible even at 2$\mu$m and is  driving a powerful massive outflow.
For instance, in the best  studied example of IRAS20126+4104, while  a
mass   of  7~\msun  is  estimated    by  using  the observed Keplerain
velocities, the luminosity derived from a spectral energy distribution
curve ($\sim$10$^4$~\lsun)  is an order of  magnitude higher than that
expected   from      a     single  7~\msun  star        (Cesaroni   et
al.~\cite{cesa05}). Further, the mass accretion rates derived from the
above  studies are in   the range of  10$^{-4}$~\msun~  to as  high as
10$^{-2}$~\msun  (Beltran  et al. \cite{beltran04}; \cite{beltran05}).
The observational evidences  consistently   suggest a scenario   where
intermediate mass objects  are  surrounded by massive  envelopes  with
sufficient accretion rates to form  a massive star through continuing
accretion. It should  be noted that contrary  to this scenario, in low
mass star formation, the  envelope mass is  greater than the protostar
mass only in Class 0 phase (M$_{env}$ $>$ M$_{*}$)  and during Class I
or II phases M$_{env}$ $<$ M$_{*}$.

\begin{table*}[t]
 \caption[]{Properties of embedded clusters associated with candidate HMPOs}   
  \begin{tabular}{@{}llrrrrrrrrr@{}}  
  \hline
  \hline
 EC$^{\mathrm{a}}$ & IRAS & d & L & \multicolumn{2}{c}{Cluster Members} &
\multicolumn{2}{c}{Effec Radius} & A$_k$ & M$_{\it stellar}$ & Outflow$^{\mathrm{b}}$ \\
 No. & Name & kpc & Log L$_{\odot}$ & mode+2$\sigma$ & true & \arcsec & pc  &
mag & M$_{\odot}$ & \\
\hline
   1* & 00117+6412 &    1.8 &   3.1 &  26 &  13 &  66  &  0.57 &  0.3 & 18 & Y \\
   2* & 00420+5530 &    4.3 &   4.1 &  71 &  38 &  96  &   3.6 &  0.6 & 380 & Y \\  
   3* & 03211+5446 &   4.54 &   4.5 & 131 &  41 &  101 &  2.21 &  0.6 & 471 & - \\ 
   4* & 04034+5116 &   3.98 &   4.2 &  50 &  16 &  112 &  2.16 &  0.3 & 100 & - \\ 
   5 & 05137+3919 &   11.5 &   4.6 &  33 &   7 &  116 &  6.08 &  0.3 & 1132 & Y \\  
   6* & 05168+3634 &   6.08 &   4.4 &  48 &  18 &  115 &  3.37 &  0.4 & 355 & Y \\
   7 & 05274+3345 &   1.55 &   3.6 &  48 &  29 &  103 &  0.77 &  0.5 & 38 & Y  \\
   8 & 05345+3157 &    2.6 &   3.6 &  95 &  36 &  126 &  1.09 &  0.6 & 122 & Y \\  
   9 & 05358+3543 &    1.8 &   3.8 &  53 &  30 &  130 &  1.13 &  0.5 & 48 & Y \\
  10* & 05373+2349 &    2.4 &   3.3 &  11 &   7 &   65 &  0.88 &  0.5 & 18 & Y \\  
  11* & 05382+3547 &   16.5 &   5.7 &  98 &  60 &  152 & 18.76 &  0.8 & - & - \\  
  12 & 05490+2658 &    2.1 &   3.5 &  95 &  43 &  122 &  1.23 &  0.6 & 94 & Y \\ 
  13 & 05553+1631 &    2.5 &   3.8 &  80 &  22 &  104 &  1.24 &  0.4 & 56 & Y \\  
  14 & 06056+2131 &    1.5 &   3.8 & 132 &  74 &  132 &  0.95 &  0.8 & 111 & Y \\   
  15 & 06061+2151 &    2.0 &   1.4 & 105 &  49 &  145 &  0.07 &  1.0 & 144 & -\\  
  16* & 06063+2040 &   4.52 &   4.9 & 202 & 114 &  121 &  2.64 &  0.7 & 1474 & -\\  
  17 & 06068+2030 &    1.5 &   4.7 &  68 &  34 &   96 &  0.69 &  0.6 & 45 & - \\
  18 & 06103+1523 &   4.3 &  4.63 & 50 &   24 &  118 &  2.64 &  0.3 & 211 & - \\  
  19 & 06104+1524 &   4.7 &  4.68 & 50 &   24 &  124 &  2.81 &  0.3 & 217 & - \\
  20* & 06105+1756 &  4.5 &  3.38 & 70 &   35 &  114 &  1.87 &   1. & 314 & - \\
  21 & 06155+2319 &    1.6 &   1.3 & 127 &  71 &  132 &  1.02 &  0.6 & 102 & - \\  
  22 & 06291+0421 &   1.96 &  1.96 &  38 &  20 &  113 &  1.07 &  0.5 & 36 & - \\  
  23 & 06308+0402 &   2.02 &   3.9 &  47 &  17 &   96 &  0.94 &  0.6 & 34 & Y \\  
  24* & 06382+0939 &   0.76 &   2.2 &  72 &  38 &  125 &  0.46 &  0.5 & 28 & - \\  
  25 & 06584-0852 &   4.48 &   3.9 &  32 &  16 &  108 &  2.33 &  0.4 & 145 & Y \\  
  26 & 20028+2903 &   1.55 &   3.7 &  37 &  10 &   72 &  1.29 &  0.7 & 15 & - \\  
  27 & 20050+2720 &   0.73 &   2.6 &  57 &  28 &  105 &  0.37 &  0.7 & 21 & Y \\   
  28* & 20081+2720 &    0.7 &   2.5 &  36 &  12 &   78 &  0.26 &   1.0 & 9 & - \\  
  29* & 20205+3948 &    4.5 &   4.5 &  44 &  17 &   74 &  1.61 &  0.6 & 190 & -\\  
  30 & 20286+4105 &   3.72 &   3.6 &  44 &   7 &   58 &  1.00 &  0.8 & 63 & Y \\  
  31* & 20319+3958 &    1.6 &   3.8 &  32 &   9 &   68 &  0.53 &   1. & 17 & - \\  
  32* & 20406+4555 &  11.9 &   5.1 &  64 &   5 &   28 &  0.69 &  0.5 & 1766 & - \\  
  33* & 20444+4629 &   2.42 &   3.5 &  81 &  17 &   57 &  0.96 &  0.3 & 37 & N \\
  34* & 21046+5110 &   0.59 &   2.4 &  62 &  24 &   85 &  0.99 &  0.6 & 17 & - \\  
  35* & 21078+5211 &   1.49 &   4.1 & 113 &  50 &  113 &  0.81 &  0.6 & 66 & Y \\  
  36* & 21202+5157 &   6.78 &   4.8 &  44 &   3 &   84 &  2.77 &  0.5 & 98 & - \\
  37* & 21526+5728 &   8.11 &   4.4 &  24 &   4 &   53 &  1.12 &  0.4 & 202 & - \\ 
  38* & 21548+5747 &    7.1 &   4.4 &  15 &   5 &   38 &   1.3 &  0.3 & 140 & - \\ 
  39 & 22134+5834 &    2.6 &   4.1 &  52 &  26 &  104 &  1.31 &  0.4 & 72 & Y \\  
  40* & 22147+5948 &   7.29 &   4.4 &  59 &   5 &  106 &  2.26 &  0.4 & 178 & - \\  
  41* & 22172+5549 &   2.87 &   3.7 & 171 &  35 &  126 &  1.74 &  0.3 & 109 & Y \\ 
  42* & 22272+6358 &   1.23 &   3.3 &  23 &  11 &   74 &  0.44 &  0.4 & 11 & - \\  
  43* & 22305+5803 &    5.4 &   4.1 &  36 &   3 &   83 &  2.16 &  0.4 & 43 & Y \\ 
  44* & 22506+5944 &    5.7 &   4.3 &  38 &  15 &   91 &  2.52 &  0.5 & 280 & Y \\ 
  45* & 22551+6221 &    0.7 &   3.2 & 229 & 129 &  152 &  0.51 &  0.8 & 98 & - \\  
  46* & 22570+5912 &    5.1 &   4.7 &  96 &  25 &   82 &  2.01 &  0.3 & 270 & Y \\ 
  47* & 23026+5948 &   5.76 &   4.2 &  59 &  22 &   86 &  2.39 &  0.3 & 331 & N \\ 
  48 & 23033+5951 &    3.5 &    4. &  10 &   3 &   74 &  1.26 &  0.3 & 14 & Y \\ 
  49* & 23146+5954 &   4.43 &   4.4 &  56 &  19 &  119 &  2.55 &  0.3 & 153 & - \\ 
  50* & 23151+5912 &    5.7 &    5. &  27 &   2 &   97 &  2.69 &  0.4 & 33 & Y \\ 
  51* & 23314+6033 &   2.78 &    4. &  176 & 61 &  126 &  1.69 &  0.6 & 239& Y \\ 
  52* & 23448+6010 &   2.02 &   3.4 &  64 &  38 &   89 &  0.86 &  0.5 & 66 & -\\
  53* & 23507+6230 &   4.28 &   4.1 &  23 &   8 &   88 &  1.81 &  0.4 & 66 & -\\   
  54* & 23545+6508 &    0.8 &   3.6 &  92 &  39 &  145 &  0.56 &  0.7 & 31 & -\\ \hline
\end{tabular}
\begin{list}{}{}
\item[$^{\mathrm{a}}$] * denotes newly detected embedded clusters. The remaining clusters were known previously as listed below\\
(5)Ishii et al. \cite{ishii02};(7,12,17,21,23) Carpenter et al. \cite{car93}; (8)Chen et al. \cite{chen99}; (18,26,30) Dutra \& Bica \cite{db01}; (9) Porras et al.\cite{porras00}; (13,14,15) Carpenter et al. \cite{car95}; (22) Phelps \& Lada \cite{pl97} ; (25) Ivanov et al. \cite{ivanov02}; (39,48) Kumar et al. \cite{kumar02}\\
\item[$^{\mathrm{b}}$]Outflow identifications from Beuther et al.\cite{beu02b}, Zhang et al \cite{zhang05} and Wu et al. \cite{wu04} \\
\end{list}
\end{table*}

\section{Summary and Conclusions}

We conducted a systematic search for clustering around 217 candidate
HMPOs chosen from the combined lists of MBCP96 and Sri02. We used the
2MASS GATOR database and the technique of producing 
stellar surface density contours to detect clusters. We also searched
for near-infrared counterparts of the 1.2\,mm dust continuum peaks
associated with all candidate HMPOs.

1) We find 54 embedded clusters associated with 217 candidate HMPOs
indicating a 25\% cluster detection rate. All targets lying in
the Galactic mid-plane did not show any clusters, and we attribute
this to the insensitivity of the 2MASS data to probe into the large
extinctions in the Galactic mid-plane region. 
The detection rate for targets away from the mid-plane is 60\%.

2) We estimate the mass of each cluster associated with massive
protostellar candidates and find that the embedded cluster mass
distribution function is similar to that found in a sample of
all embedded clusters withing 2 kpc of the Sun.

3) Approximately equal numbers of clusters associated with
massive protostellar candidates are found to have
hierarchical as centrally condensed structures.

4) In about half of the detected sample, the cluster peaks and the
IRAS/mm peaks coincide very well indicating a positive identification
of the massive protostar with the infrared visible embedded
clusters. This fraction of clusters also display the highest densities
and more circular morphology in the entire sample.

5) One  hypothesis of the formation  of   massive stars by continuing
accretion is that the younger stars will not  be associated with radio
free-free  emission from HII regions. We  find that the near IR colors
of  sources  near the dust   continuum  peaks and  that   do not  show
free-free emission  are redder than  those  that are  associated  with
ultra-compact HII regions.

6) The data are consistent with the hypothesis that massive stars
form by continuing accretion, but the data do not discriminate
against the hypothesis that massive stars form by the collisions
of lower mass stars.

\begin{acknowledgements}

We are grateful to Alessandro Chieffi for computing the evolutionary
tracks shown in Fig.\,5. We also thank an anonymous referee for
valuable suggestions that significantly improved the presentation of
this paper. This publication makes use of data products from the Two
Micron All Sky Survey, which is a joint project of the University of
Massachusetts and the Infrared Processing and Analysis
Center/California Institute of Technology, funded by the National
Aeronautics and Space Administration and the National Science
Foundation.  This work was supported by grants POCTI/1999/FIS/34549
and POCTI/CFE-AST/55691/2004 approved by FCT and POCTI, with funds
from the European Community programme FEDER.

\end{acknowledgements}

\end{document}